# Synthesis of single-cycle pulses based on a Yb:KGW laser amplifier

## Z. PI,[1] H. Y. KIM,[1] AND E. GOULIELMAKIS[1*]

[1]*Institute of Physics, University of Rostock, 18059 Rostock, Germany*
*Corresponding author: e.goulielmakis@uni-rostock.de*

Yb:KGW lasers have been gaining increasing importance in ultrafast and strong-field physics, but their compression to the single- and sub- cycle regime remains a challenge. Here, we utilized a dual hollow-core fiber module for the spectral broadening of Yb:KGW pulses to more than 3.5 optical octaves, from the short-wave infrared (~1900 nm) to vacuum ultraviolet (~170 nm). A light-field-synthesizer compressed a large portion of this bandwidth (380-1000 nm) to single-cycle pulses for the first time, based on Yb:KGW technology. Our work opens the door to attaining new regimes of control and temporal compression of light pulses and their advanced applications in ultrafast spectroscopy.

## 1. INTRODUCTION

Lasers based on Ti:Sapphire crystals have long been at the forefront of cutting-edge light technology, enabling the generation of single [1] sub-cycle [2-3], and optical attosecond pulses [4] in the visible and adjacent ranges of the electromagnetic spectrum. Such pulses have played a pivotal role in the development of strong field physics in atoms, molecules, and solids as well as the advancement of attosecond science [5]. Developments of the Yb:KGW laser technology over the past decade have opened the promise for a new generation of ultrafast lasers with scalable energy and repetition rates, which are soon anticipated to take the lead in advancing ultrafast science to the next level. Nevertheless, compression of their pulses to the single-cycle regime has remained beyond reach. Key reasons include the relatively long duration of these pulses (~170 fs) delivered by these systems compared to their Ti:Sapphire counter parts, and the decreased nonlinearity of gas media at infrared carrier frequencies. As a result, the compression factor required (~70) to attain the single-cycle regime is challenging. Efforts over the last decade using nonlinear methods have made significant steps towards the ultrafast compression of these pulses generated by Yb:KGW amplifiers [6-20]. Utilization of cascade hollow-core fiber modules allowed further compression to pulse shorter than two cycles of their carrier frequency [21]. Here we show that the extension of this technology with pressurized Neon filled hollow-core fibers and subsequent compression of the generated supercontinuum pulses via light field synthesis allow the advancement of this technology to the single-cycle regime and beyond.

## 2. EXPERIMENTAL SETUP

In our experiments, pulses from a Yb:KGW laser system (Pharos, Light Conversion) with a duration of ~170 fs, carried at a central wavelength of ~1030 nm, and energy of ~1 mJ per pulse at 6 kHz were used. The pulses were coupled by a fused silica lens (f = 75 cm) into the entrance of a hollow-core silica capillary (inner diameter ~250 µm, length 0.55 m) housed in a high-pressure (~18 bars) Ne-filled chamber as shown in Fig. 1(a). Antireflection-coated fused-silica entrance and exit windows with a thickness of ~2 mm minimized nonlinear effects on the incoming and outgoing laser beam through the module. Moreover, to further mitigate undesirable nonlinearities, the focusing lens was placed only a few centimeters away from the entrance window of the host chamber to allow a large beam spot and correspondingly low intensity. Through a viewport, located at the top of the host chamber (Fig. 1(a)), a CCD camera monitored the glass capillary entrance, allowing reproducible and damage-free day-to-day coupling of the laser beam into the hollow-core fiber. The pulses emerging from the first stage were compressed by reflections off six dispersive mirrors (4 on HD120 and 2 on PC147, UltraFast Innovations), yielding a total group delay dispersion of –920 fs$^2$ while fine adjustments were possible by pair of thin Brewster-angled silica wedges.

Fig. 2(a) presents a FROG spectrogram of the pulses exiting the fiber module, recorded by an all-reflective Transient Grating setup (TG-FROG) [22,23], schematically illustrated in Fig. 2(b). An excellent reconstruction quality is revealed by the data in Fig. 2(b), which is further verified by the evaluated FROG error of 0.0025 (1024 × 1024 matrix). Fig. 2(c) compares the measured and retrieved spectra of the pulses, along with their spectral phase. The temporal intensity pulse profile and phase are also shown in Fig. 2(d). The FWHM duration of the pulses was ~20.5 fs, slightly longer than the Fourier-limited duration of ~18.6 fs, as evaluated by the measured spectral bandwidth. As the pulses uniformly broadened towards longer and shorter wavelengths, the carrier wavelength of the incoming pulse was retained. The laser power at the exit of the compressor was approximately 4.7 W corresponding to a 90% efficiency.

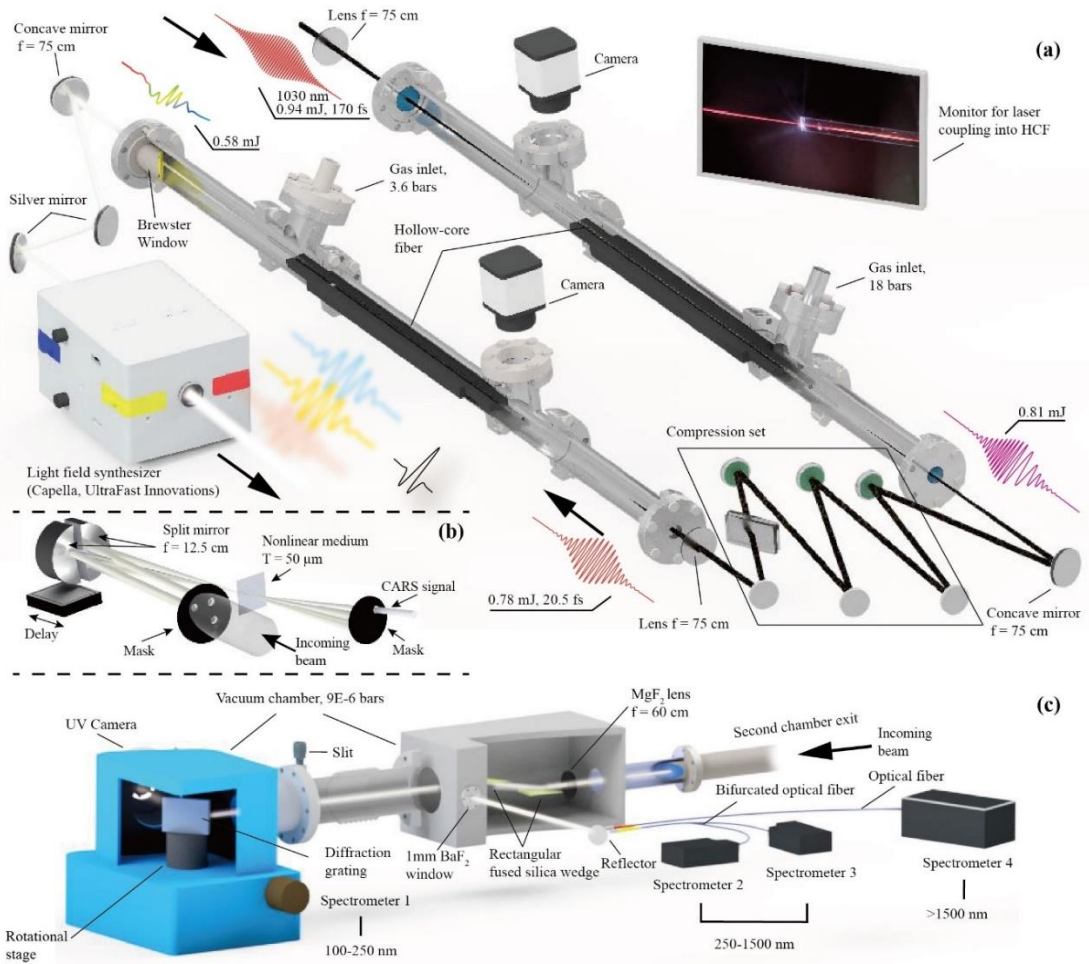

**Fig. 1**. (a) Two-stage, cascaded, hollow-core fiber (HCF)-based pulse broadening module. (b) Schematic of the transient grating frequency-resolved optical gating (TG-FROG) apparatus. (c) The experimental setup for measuring the spectrum of the generated supercontinuum pulses.

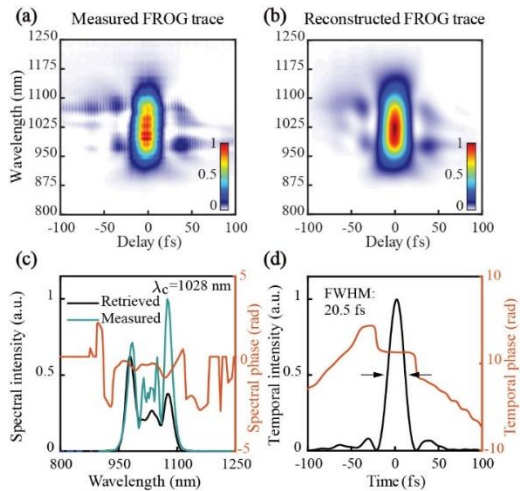

**Fig. 2**. Temporal characterization of the pulses at the exit of the first compression stage. (a) Measured TG-FROG spectrogram. (b) Reconstructed spectrogram. (c) Measured spectrum (cyan) and retrieved spectrum (black). The orange curve denotes the retrieved spectral phase and $\lambda_c$ is the carrier wavelength of the compressed pulse. (d) Retrieved temporal intensity profile (black) and temporal phase (orange) of the generated pulses.

In the next step, a second fused silica lens (f = 75 cm) coupled the compressed pulses into the entrance of a second hollow-core fiber (Fig. 1(a)) whose properties are identical to those of the first stage with fiber length of 0.55 m. The Ne pressure in the host chamber was held at ~3.6 bars. To allow measurements of light spectra down to the vacuum ultraviolet range <200 nm, where air absorption is significant, a vacuum chamber was directly attached to the exit of the host chamber and was pumped to a background pressure of 9E–6 bar, as illustrated in Fig. 1(c). The two chambers were separated by a thin (2 mm) $CaF_2$ window installed at the one end of a flexible connecting bellow. The latter allowed quasi-free tilting of the fiber-chamber permitting optimization of the beam mode and energy output. The generated supercontinuum pulses were collimated by a $MgF_2$ lens (f = 60 cm) placed downstream from the source and were reflected off a pair of wedges to attain a spectrally uniform

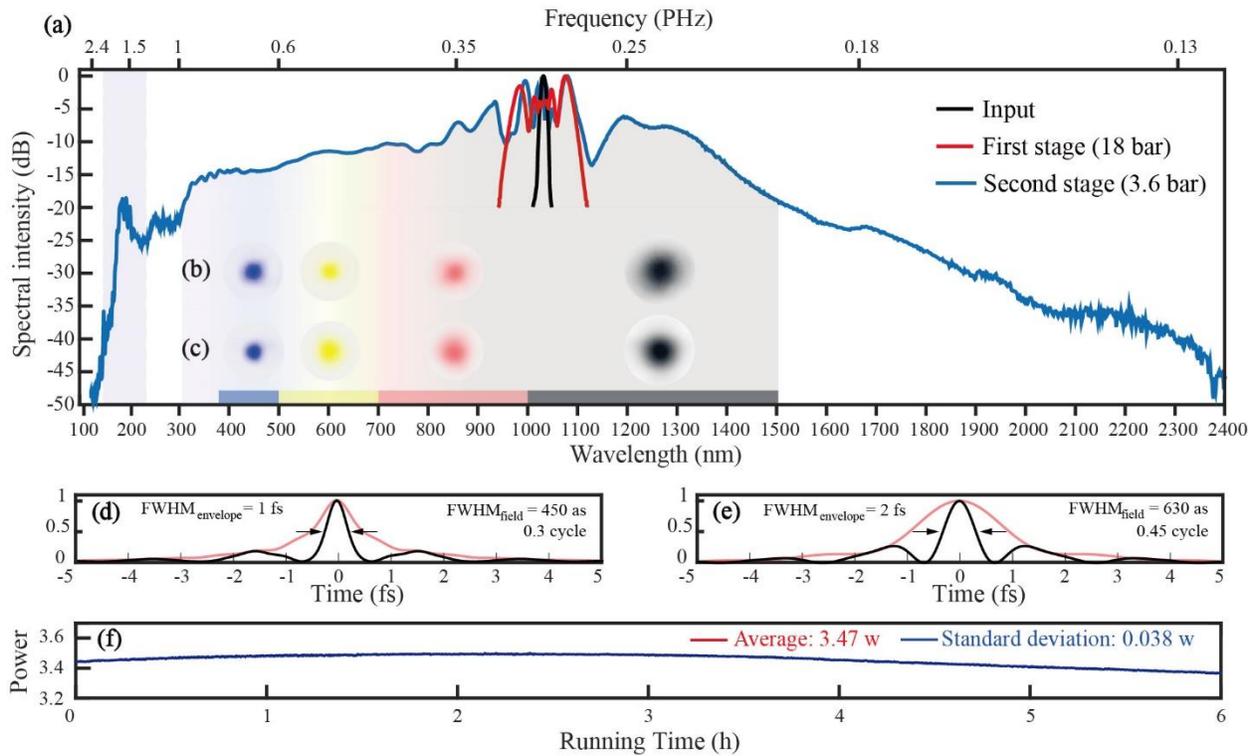

**Fig. 3.** Broadband supercontinuum generation. (a) Spectrum of the pulses at the exit of the laser (black line), the first stage post-compression (red line). Color block at the wavelength axis denote the spectrum above 20 dB range. (b)-(c) Beam profile at near-field and far-field respectively. (d) Evaluated instantaneous intensity profile (red line) and intensity envelope (black line) of the Fourier-limited pulse associated with the entire supercontinuum at the exit of the second hollow-core fiber. (e) is same as (d) but within an intensity dynamic range of ~20 dB. (f) Power stability at the exit of second hollow-core fiber module over a time interval of 6 hours.

attenuation of power before they were routed to the entrance of the spectrometer.

Four spectrometers were employed for the measurement of the spectra emanating from the second fiber module. A Macpherson vacuum spectrometer covered the range from 100 nm to 250 nm, an HR4000, Ocean Optics for the range 250 nm to 900 nm, a NIRQuest, Ocean Optics for the range 900 nm to 1500 nm, and a scanning spectrometer WaveScan MIR for wavelengths above 1500 nm. The spectral intensity calibration, in the range 150 nm to 250 nm, was performed using a deuterium source (Macpherson 632), while for wavelengths >250 nm a tungsten source (DH-3P-CAL, Ocean Optics) was used.

## 3. SPECTRAL DOMAIN STUDIES

Fig. 3(a) (blue line) shows spectra of the generated supercontinuum pulses as measured at the exit of the second fiber module. The initial pulse spectra of laser system (black line) and these at the exit of the first stage (red line) are also included in the plot for comparison. The spectra extends from ~123 nm to ~2400 nm over a dynamic range of ~50 dB corresponding to 4.27 optical octaves from ~170 nm to ~1889 nm at ~30 dB corresponding to 3.4 octaves. The formation of a peak at the UV part of the spectrum is compatible with these predicted in recent studies of soliton generation and propagation in hollow-core laser waveguides [24,25]. We also verified that the supercontinuum light source has adequate power stability to enable future, high precision strong-field and attosecond experiments. To this end, we monitored the power of the supercontinuum pulses at the exit of the host chamber. Fig. 3(f) shows measurements recorded over a period of ~6-hour. The averaged power remained nearly constant at 3.47 W, with a standard deviation of ~0.038 W. This corresponds to a long-term stability better than 1.1%. It is essential to add that no beam-pointing stabilization was required at any stage of the experimental setup to attain this stability.

The potential of these broad supercontinua for the synthesis of ultrashort pulses is highlighted in Fig. 3(d) and (e), which plots the Fourier-limited waveforms and corresponding pulse intensity envelopes based on the measured spectra in Fig. 3(a). Indeed, the entire spectral range could support the synthesis of a pulse whose intensity envelope would be confined to ~1 fs, containing 0.3 field cycle. Moreover, the instantaneous intensity profile (shown in black) is confined to merely 450 attosecond (as). By limiting the bandwidth to a more practical dynamic range of the spectral intensity (~20 dB) the corresponding bandwidth could support pulses of ~2 fs or ~0.45 field cycle, respectively. Given these estimations, the broadening factor of the entire dual-stage setup could also be evaluated to approximately 85, which is, well beyond the requirements of the single-cycle.

## 4. CHANNELS TEMPORAL COMPRESSION AND LIGHT FIELD SYNTHESIS

To synthesize single-cycle transients, we compressed a large fraction of the generated supercontinua (380-1000 nm) using a light-field synthesizer. The technology of light-field synthesis is detailed in previous works [3,4,26,27]. Briefly, the spectral range from ~380 nm to 1000 nm was divided into three spectral bands (hereafter: channels): $Ch_{UV-VIS}$ (380-500 nm), $Ch_{VIS}$ (500-700 nm), and $Ch_{NIR-A}$ (700-1000 nm) whose bandwidth is sufficiently narrow and thus compressible by dispersive mirror technology. Splitting of the original pulses into individual channels was possible by broadband dispersive beam splitters designed for the entire spectral range [28,29]. The total dispersion that was required for the compression of the pulses in each channel was ~7 × 75 $fs^2$ for the channel $Ch_{UV-VIS}$, ~5 × 70 $fs^2$, for the channel $Ch_{VIS}$, and ~4 × 70 $fs^2$ for the channel $Ch_{NIR-A}$, respectively, while a pair of thin fused-silica wedges (angle of 2° 48') in each channel allowed the fine-tuning of the dispersion of the corresponding pulses.

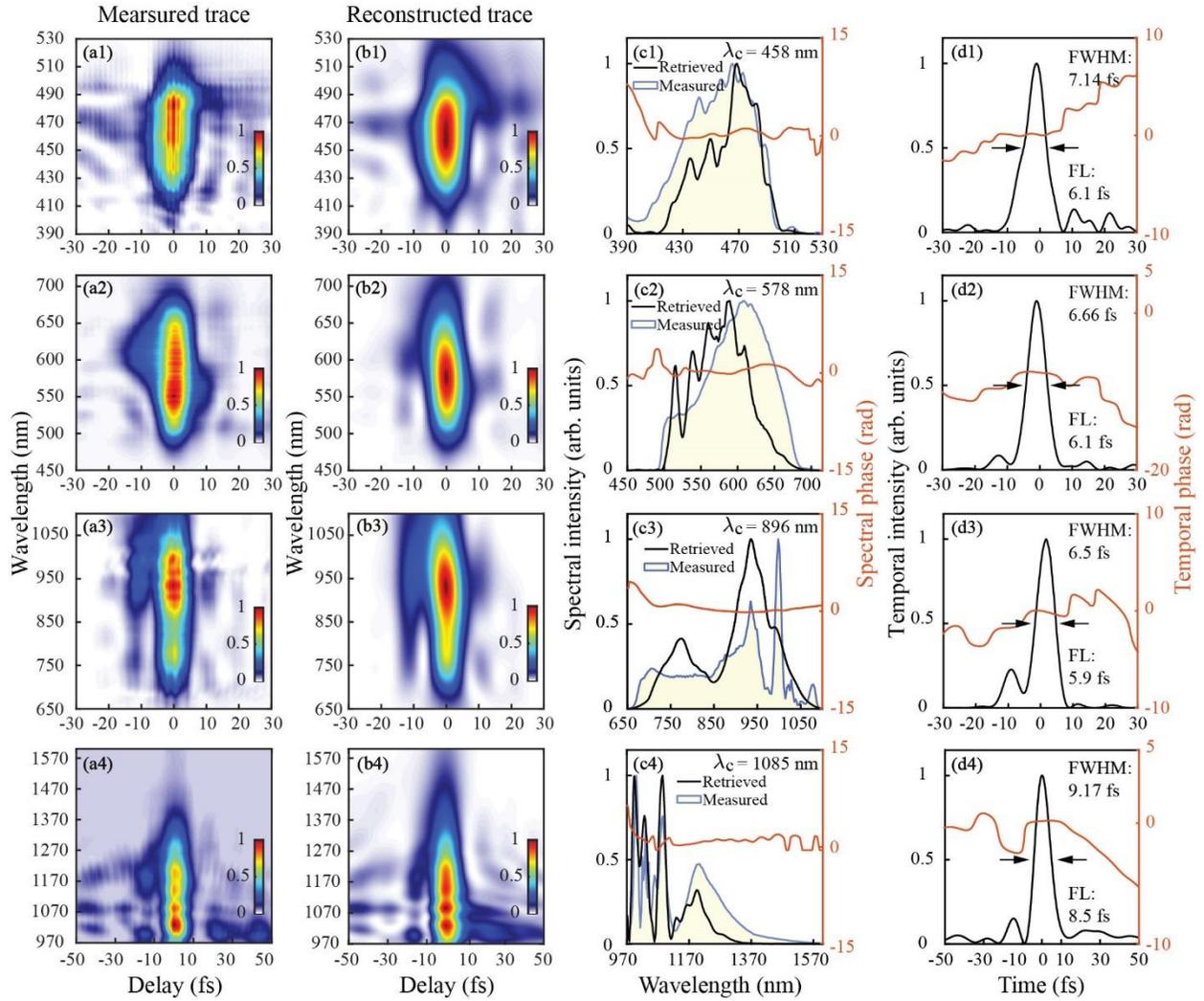

**Fig. 4**. Characterization of channel pulses. a (1-4) Experimentally measured TG-FROG trace. b (1-4) Reconstructed TG-FROG trace. c (1-4) Measured spectrum (yellow area covered by blue solid line). Retrieved spectrum (black) and its spectral phase (orange), where $\lambda_c$ denotes the carrier wavelength. d (1-4) Temporal intensity profile (black) and temporal phase (orange) of the retrieved pulse. The four spectral channels are as follows: (a-d) 1 $Ch_{UV-VIS}$ (380-500 nm), (a-d) 2 $Ch_{VIS}$ (500-700 nm), (a-d) 3 $Ch_{NIR-A}$ (700-1000 nm), and (a-d) 4 $Ch_{NIR-B}$ (1000-1500 nm). (d) Retrieved temporal intensity profile (black) and temporal phase (orange).

Fig. 4 summarizes TG-FROG measurements of pulses in each channel of the field synthesizer. In $Ch_{UV-VIS}$, the pulses had a duration (FWHM) of $\tau_{UV-VIS}$ = 7.1 fs corresponding to 4.6 field cycles of their carrier wavelength (458 nm). In $Ch_{VIS}$, the duration of the pulses was $\tau_{VIS}$ = 6.6 fs corresponding to 3.4 field cycles of 578 nm, while for the near-infrared channel, the pulse duration was $\tau_{NIR-A}$ = 6.5 fs corresponding to 2.1 field cycles at 896 nm. All measured pulses durations were close to their Fourier-limited duration (FWHM): $\tau_{UV-VIS}$ = 6.1 fs, $\tau_{VIS}$ = 6.1 fs, and $\tau_{NIR-A}$ = 5.9 fs, as evaluated by the recorded spectra in each channel (Fig. 4(c1) to Fig. 4(c3)).

Although not an integral part of the synthesizer at this stage of our research, we have attempted compression of the pulses in the short-wave infrared (1000-1500 nm). Pulses in this range were isolated by a thin long-pass filter (Schott, RG1000) and were compressed by two bounces off dispersive mirrors PC1816 (-70 $fs^2$ per reflection, UltraFast Innovations) and a pair of wedges. The temporal characterizations of these pulses are summarized in Fig. 3 (a-d)4 revealed a duration of $\tau_{NIR-B}$ = 9.17 fs or 2.53 field cycles centered at 1085 nm. The power at each optical channel was $Ch_{UV-VIS}$: 52 mW, $Ch_{VIS}$: 163 mW, $Ch_{NIR-A}$: 512 mW, and $Ch_{NIR-B}$: 1.1 W. Since high-quality spatial characteristics of the generated supercontinuum pulses is essential for applications of this novel source, we performed beam profile measurements, at both near- and far-field (focus length f = 30 cm), using a commercial profiler. Measurements were conducted for individual spectral channels, as no single-beam monitoring device could offer sensitivity to the entire spectral range. The near-field beam diameters for each of the above ranges were 2 mm, 2 mm, 3 mm, and 3.9 mm, as measured at the $1/e^2$ of their corresponding beam profiles. At the far field, the focal diameters of the same spectral channels were evaluated as $d_{UV-}$ $v_{IS}$ = 85 μm, $d_{VIS}$ = 121 μm, $d_{NIR-A}$ = 165 μm, and $d_{NIR-B}$ = 157 μm, respectively. These results, along with the corresponding profiles acquired, are shown as insets in Fig. 3(b)-(c), suggesting an excellent beam quality of the emerging supercontinua.

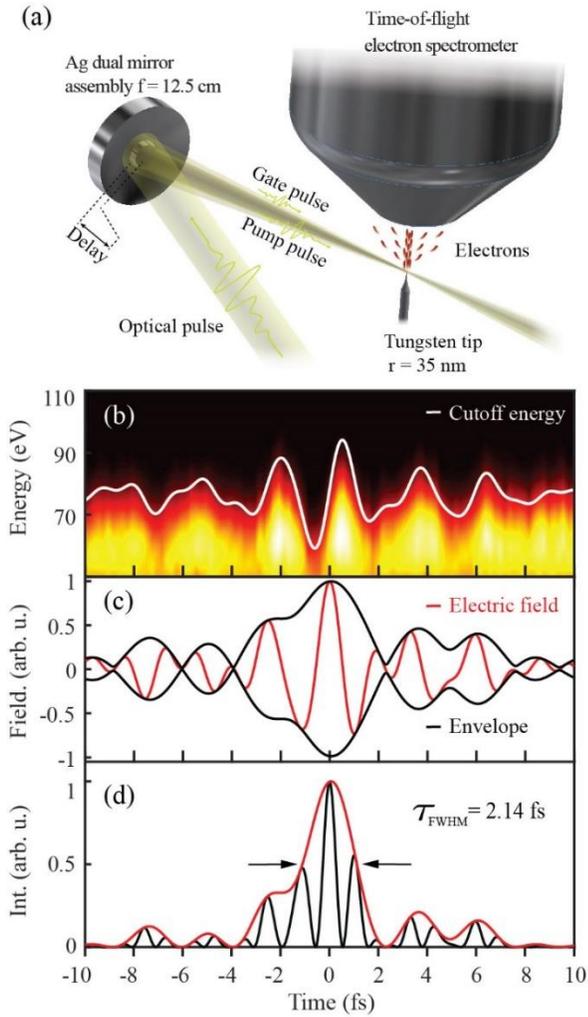

**Fig. 5.** HAS characterization of a single-cycle pulse. (a) Experimental setup of the HAS apparatus. (b) A HAS spectrogram recorded for the synthesized pulse (false color). White line denotes the variation of the energy cutoff of the electron spectra as a function of the delay between pump and gate pulses. (c) Electric field (red line) and field envelope (black lines) as derived from the data in (b). (d) Instantaneous intensity profile (black line) and intensity envelope (red line) of the synthesized pulse whose FHWM is confined to ~2.14 fs (0.85 field cycle).

To synthesize a single-cycle pulse, we involved three channels ($Ch_{UV-VIS}$, $Ch_{VIS}$, and $Ch_{NIR-A}$). Accurate measurements of the synthesized pulse were performed using the recently developed technique of Homochromatic Attosecond Streaking (HAS). The synthesized pulse was reflected off a dual concave mirror assembly, which consists of two concentric and silver-coated, inner- and outer- mirrors, as depicted in Fig. 5(a). In this way, the optical beam was divided into an intense inner beam (pump pulse) and a weak outer beam (gate pulse) that were in turn focused onto an electrically grounded tungsten nanotip (apex radius of ~35 nm). A piezo translation stage on which the inner mirror was attached controls the delay between pump and gate pulses with attosecond precision. HAS spectrograms were composed by recording electron spectra as a function of the delay between pump and gate pulses using a time-of-flight spectrometer whose entrance is located ~2 mm above the tungsten nanotip. Fig. 5(b) shows the HAS spectrogram recorded for our synthesized pulse. The white curve in Fig. 5(b) tracks the variation of the electron cutoff energy as a function of the delay. The HAS vector potential is directly linked to the vector potential $A_g(t)$. This can be best expressed in the Fourier domain, $\tilde{A}_{HAS}(\omega) = \tilde{A}_g(\omega)\tilde{g}(\omega)$, where the multiplier $\tilde{g}(\omega)$ is given by $\tilde{g}(\omega) = \left[2 - \frac{1}{\omega \Delta t}(e^{-i\omega \Delta t} - 1)\right]$ [30,31] while the electric field of the synthesized transient is related to the vector potential as $E_g(t) = -dA_g(t)/dt$.

Fig. 5(c) shows the retrieved electric field of the synthesized pulse and its evaluated envelope. While Fig. 5(d) shows instantaneous intensity of the synthesized pulse and its intensity envelope (black line) whose FWHM is confined to $\tau$ = 2.14 fs. The evaluated centroid wavelength is 736 nm, which corresponds to an optical period of 2.35 fs. This implies that the synthesized pulse contains 0.85 optical cycles within the FWHM of its intensity profile, and it is somewhat longer than the Fourier-limited duration associated with its bandwidth ~1.95 fs. The synthesized pulse energy at the exit of the device was >120 μJ.

## 5. CONCLUSION

We have demonstrated, for the first time, the generation of powerful coherent supercontinuum pulses based on Yb:KGW laser amplifiers that extend over more than ~3.5 octaves, *i.e.*, covering the range from the short-wave infrared (~1500 nm) to the vacuum ultraviolet (~150 nm). To this end, cascade broadening of the original pulses in a dual stage hollow-core fiber setup has been an essential strategy. Such supercontinua could, in principle, support the synthesis of optical attosecond pulses a tool earlier possible only by Ti:Sapphire amplifiers. We have shown that light field synthesis over approximately 1.5 octaves of the generated supercontinua allows, for the first time, the generation of single-cycle pulses from a Yb:KGW amplifier. Combination of these pulses with those in the short-wave infrared generated by the same source would allow a broad range of ultrafast studies in low band gap and correlated materials as well as strong field experiments in all phases of matter.

**Disclosures**. The authors declare no conflicts of interest.

**Data availability**. Data underlying the results presented in this paper is not publicly available at this time but may be obtained from the authors upon reasonable request.